\documentclass[final]{siamltex}
\usepackage[dvips]{graphicx}
\usepackage{alltt}
\usepackage{fancyheadings}
\usepackage{amsmath}
\author{Olivier Cessenat\thanks{CEA/DAM/CESTA, F-33114 Le Barp, France.}}
\title{Sophie, an FDTD code on the way to multicore, getting rid of the memory bandwidth bottleneck better using cache.}
\date{\today.}
\begin{document}
\maketitle
\begin{abstract}
FDTD codes, such as Sophie developed at CEA/DAM, no longer take
advantage of the processor's increased computing power, especially
recently with the raising multicore technology. This is rooted in the fact
that low order numerical schemes need an important memory bandwidth to
bring and store the computed fields.
The aim of this article is to present a programming method at the
software's architecture level that
improves the memory access pattern in order to reuse data in cache
instead of constantly accessing RAM memory. We will exhibit a
more than two computing time improvement in practical applications.
The target audience of this article is made of computing scientists
and of electrical engineers that develop simulation codes with no
specific knowledge in computer science or electronics.
\end{abstract}

\section{Introduction}\label{intro}
At CEA, the French Nuclear Agency, we develop yet another FDTD (Finite
Difference in Time Domain) code, called Sophie, for electro-magnetics
simulations of the Laser MegaJoule (LMJ) device.
To achieve realistic three-dimensional simulations, we need to be able to make
computations on tens of billions of cells. This challenging goal not only
demands a massively parallel machine (Tera 10, see \verb+www.top500.org+), but also a very efficient
implementation to reduce the computation time on an always very busy
High Performance Computing centre. Current major LMJ simulations cost
roughly $500$ Euros in electric power consumption. Improving the
computing's efficiency is an economic issue.

FDTD codes date back to the late sixties with their numerical scheme
first presented in \cite{Yee}, becoming very popular in the eighties
with the appearance of the first Cray vector super-computers that made
three dimensional simulations affordable \cite{Taflove}.
Those codes were very fast due to a perfect vectorization level,
particularly efficient on large problems \cite{Gems}.

Nowadays, very large simulations are performed using many mass market
personal computers (PCs) components. On those hardware platforms, FDTD codes, as well as all
low order solvers, are not able to take advantage of the still
up-to-date Moore's law that applies to the processors peak's computing
power. This is rooted in the fact that low order solvers need an
important bytes to flops ratio (the program balance as described in
\cite{DingKen}), typically $4$ for an FDTD code in single precision.
Ever since thirty years, whereas CPU speed doubles in average
every $18$ months, memory bandwidth takes around $33$ months to double
according to \cite{Ding}. Moore's law validity in the future will
probably rely on increasing the number of cores per processor whereas
there seems to be no plan to make cores manage the memory accesses
independently.

Since the mid-nineties, memory bandwidth bottleneck is seen as the
major obstacle to scientific computing generally made of sparse matrix to
vector products where modern processors spend more than $80\%$ of their
time waiting for data \cite{Gems}. This is visible to the scientific
kernels included in the Specfp benchmarks (\verb+www.spec.org+) such
as GemsFDTD for the $3D$ FDTD.
Much work is being performed to overcome this memory bandwidth
bottleneck through very clever cache hierarchies and compiler's
optimizations (refer to \cite{DingKen} and \cite{GaoCache1} for cache
use optimizations and to \cite{GaoRegister} for register pressure in
multidimensional loops).
These optimizations provide a real benefit on many basic linear
algebra applications, but their essential weakness is that they
lack a global view of the scientific application (they are very often
limited to a procedural optimization).

Another possibility is to use specific hardwares such as FGPAs where data bandwidth
is provided in parallel as in \cite{Cell}.
But specific VLSI programming is far too complicated for computing
scientists with no knowledge in electronics.
Another drawback of specific electronics is that simulation code's lifetime
often exceeds twenty years, whereas electronic devices cannot last
more than two years. CPUs life cycle is around four years, but high
level programming languages such as FORTRAN limit the cost of porting
codes from a CPU to another.
Recently, using GPUs (Graphics Processing Units) is seen as an alternative or a
complement to general CPUs with the development of CUDA, an NVIDIA
C-styled language \cite{CUDA}. Even though NVIDIA's products lifetime
is typically two years, the CUDA language is expected to be supported
during generations of products.
Depending on the application's computing kernel's features, a speed
gain of $10$ to $50$ can be reached.
For instance, on the FDTD kernel, a factor of $10$ is
obtained in \cite{CUDA}. The drawbacks are that GPUs typically provide
more than an order of magnitude less storage than the system's RAM and that the
kernel's efficiency is masked by the other parts of the application,
making the use of a GPU inefficient if the output of the device's
computation has to be transfered into RAM through the PCI bus too
often.

On the long run however, vendor independent environments, such as HMPP
developed by ``CAPS entreprise'', aiming at providing standard directives, as with
OpenMP, seem very attractive. Instead of providing a low level API to
GPUs programming, HMPP takes advantage of the high level features related to
the program. It then generates optimized GPU code using the low level
vendor's API. It is further described in \cite{HMPP1} and \cite{HMPP2}.
It also provides a resource manager which is very useful
in an SMP multi-GPU environment.

When optimization both at the hardware and at the compiler's level
fail, computing scientists need to review their algorithms so that
maximizing the processor's data cache use (and, if possible, data cache
hierarchies) is taken into account at the software's design step
preceding the implementation. This is not the first time that
numerical schemes or computing algorithms must be reviewed to better
suit the available underlying computing platforms. The Cray vector computers in
the eighties gave a boost to linear algebra and the standard Galerkin
methods such as the Finite Elements and the FDTD. The parallel
machines era in the nineties demanded breakthroughs in the field of
implicit solvers with domain decomposition methods. Today, high order
methods with more flops per byte such as the Discontinuous Galerkin
Method should address the broadening hardware's gap between CPU and memory
bandwidth \cite{DGPRF}, at least on Cartesian meshes.

The aim of this article is to present techniques at the application's
architecture level to naturally optimize the cache use as in
\cite{TUM1} and \cite{TUM2}, giving a computing scientist's point of view.
These techniques apply on todays multicore processors that have enough
cache. We believe our approach is original since we consider the
algorithm as a whole, not merely one time step. Even though we focus
on a time explicit scheme, the method can be adapted to domain
decomposition iterative implicit solvers where the iterative steps play the
role of the time steps.

Since optimization at the application level needs an insight on the
application's features, section \ref{frame} presents the basic
framework for any FDTD code. The puzzling practical experiment is a
parallel efficiency superior to $1$ on mid-sized problems.
A simplified (considering only one cache level with infinite bandwidth
to processor's registers) theoretical analysis
is presented section \ref{theory}, with major expected speed gains.
Confrontation with numerical experiments is shown section
\ref{numerexp}, first on the FDTD kernel then on the whole FDTD
production code Sophie for electro-magnetics simulations at CEA. It
demonstrates an enhancement of the computation speed by a factor two.

\section{Basic framework issues}\label{frame}
Section \ref{present} presents the basics of the FDTD scheme.
In section \ref{issue}, we explain why FDTD codes struggle against
the memory bandwidth bottleneck.
The processor's peak computing power is only available when problem is
small enough so that it's fields fit in cache as explained section \ref{cache}.
In today's era of massive parallel machines, FDTD codes bear the
advantage to be able to fit in cache when the number of computation
nodes increases, as we see section \ref{parallel}. Thus, a parallel
efficiency superior to $1$ can be estimated and actually encountered on
practical problems.

\subsection{FDTD scheme overview}\label{present}
The Finite Difference in Time Domain is the discrete scheme, leapfrog
in time and space, of the Maxwell equations in vacuum. Let $\boldsymbol\nabla\times$
denote the curl operator, $\varepsilon_0$ be the permittivity of
vacuum, $\mu_0$ the permeability of vacuum. The Maxwell equations
write:
\begin{equation}\label{ma}
\varepsilon_0\partial_t \boldsymbol{E} - \boldsymbol\nabla\times \boldsymbol{H}  = 0 \ \ \mbox{    [Maxwell-Ampere]}
\end{equation}
for the electric field evolution, and
\begin{equation}\label{mf}
\mu_0 \partial_t \boldsymbol{H} + \boldsymbol\nabla\times \boldsymbol{E}  = 0 \ \ \mbox{    [Maxwell-Faraday]}
\end{equation}
for the magnetic field evolution.

These equations can be made fully antisymmetric with variables
$\tilde{E}=\varepsilon_0 E$ and $\tilde{H}=H/c$ as follows:
\begin{equation}\label{ma2}
\partial_t \tilde{\boldsymbol{E}} - c \boldsymbol\nabla\times \tilde{\boldsymbol{H}}  = 0 
\end{equation}
\begin{equation}\label{mf2}
\partial_t \tilde{\boldsymbol{H}} + c \boldsymbol\nabla\times \tilde{\boldsymbol{E}}  = 0
\end{equation}
The Yee space discretization is the integral form of these equations
using Green's formula (refer to \cite{MC96} for the mathematical
framework) on faces of the mesh for Faraday's equation, on the edges
for Ampere's equation, as detailed in \cite{Yee}. Time discretization
is a leapfrog scheme. Thus, updating the magnetic field in direction
$x$ is (refer to \cite{Taflove} or \cite{Gems} for the other fields):
\begin{eqnarray*}
H^{n+\frac{1}{2}}_{i,j+\frac{1}{2},k+\frac{1}{2}}(x) = H^{n-\frac{1}{2}}_{i,j+\frac{1}{2},k+\frac{1}{2}}(x)
   & + \frac{c\Delta t}{\Delta z_{k+\frac{1}{2}}}
\left(E^{n}_{i,j+\frac{1}{2},k+1}(y)-E^{n}_{i,j+\frac{1}{2},k}(y)\right) \\
   & - \frac{c\Delta t}{\Delta y_{j+\frac{1}{2}}}
\left(E^{n}_{i,j+1,k+\frac{1}{2}}(z)-E^{n}_{i,j,k+\frac{1}{2}}(z)\right)
\ .
\end{eqnarray*}

The problem's numerical stability requires a perfect antisymmetry on
the discrete equations which is true when solving equations \ref{ma2}
and \ref{mf2} on a Cartesian mesh \cite{GR00}.

\subsection{FDTD codes and memory bandwidth issues}\label{issue}
Processor to memory throughput nowadays is typically 6.4 GB/s with
DDR2-800 (JEDEC standard PC2-800, see \verb+www.jedec.org+) or up to 12.8 GB/s with the latest
DDR3 modules (JEDEC standard PC3-1600, abusively referred to as PC3-12800).

In FDTD codes, the maximum number of cells that we can compute with
DDR2-800 is in practice typically 50 MC/s with 32 bits precision.
This computing power is quasi independent of the processor's frequency
and of the processor's number of cores.

This fact is well known from the Specfp 2006 benchmark with the GemsFDTD
test module from the ``Center for Parallel Computers'' (PDC) at the Royal University
of Stockholm, Sweden. In the bench description \cite{Gems}, many
numerical experiments on different hardwares (IBM, Fujitsu, SGI,
Intel) and an extensive analysis detail which is the best
implementation for each platform. Apart from the Fujitsu vector
super-computer, all machines have poor performance on large problems,
poor related to the processor's peak speed. We strongly recommend the
reader to refer to this article \cite{Gems} to better understand facts that shall not
be re-explained here.

The reader should not confuse Gems from PDC (that includes GemsFDTD)
with the FDTD leading commercial software, massively
parallel software as proven by tests on Blue Gene/L that targets
petaflopic computations \cite{ComGems}.

Let us explain why FDTD codes demand a high number of bytes to flops
ratio, considering one of the six FDTD triple loops implemented in the
standard way as given in A. Taflove's handbook \cite{Taflove} p.
$547$ following Yee's numerical scheme \cite{Yee}:
\begin{alltt}
do k=1, nz; do j=1, ny; do i=1, nx-1
  hx(i,j,k)=hx(i,j,k)+ &
  & dz(k)*(ey(i,j,k+1)-ey(i,j,k))+dy(j)*(ez(i,j+1,k)-ez(i,j,k))
end do; end do; end do
\end{alltt}
To compute \verb+hx(i,j,k)+ we need to fetch four electric field
values and the magnetic field value itself, then store the
result. This makes 6 memory accesses.

We do not deal with the geometrical coefficients \verb+dz(k)+ and
\verb+dy(j)+ since we assume they fit in registers. Concerning the
\verb+dx(i)+ array used for the computation of \verb+hy(i,j,k)+ and
\verb+hz(i,j,k)+, we assume it fits in cache. This is an important
specific point of the Cartesian programming: the sparse matrix that
updates the field is stored in one-dimensional arrays.

FDTD computations are most of the time performed in single precision,
but not always. This is why we shall consider both single and double
precision. A real float size in single precision is 4 bytes wide ($32$
bits), a double takes 8 bytes ($64$ bits).

Thus, using $6400$ MB/s memory throughput, we compute in single
precision ($4$ bytes), knowing there are 6 fields per cell and 6 data
access per field:
\begin{equation}\label{eq_ram_standard}
\frac{6400}{4\times 6\times 6} = 44
\end{equation}
million cells per second (MC/s). In double precision, we can compute 22
MC/s. Of course, this analysis does not take
into account the cache effect, which reduces the memory
requirement. This is why, in practice, even for very large domains
where cache benefits are reduced, computing speed is slightly higher.

On an Itanium II dual core processor at 1600 MHz, we can perform 4
floating point operations per cycle and per core, thus 12800 MFlops.
In the above FDTD scheme, we need to perform 6 operations per field.
We could expect to compute, knowing there are 6 fields per cell:
$$
\frac{12800}{6\times 6} = 355 \mbox{ MC/s}.
$$
Indeed, Itaniums can actually perform 4
operations per cycle when we need to compute a product then an
addition. This is not true in the FDTD algorithm where we
essentially need to compute an addition then a product. This is why
the algorithm can benefit from only half the Itanium's II amazing
computing power.

Making the ratio between the number of cells that the processor can
compute to the number of cells that memory can feed, we find that
the compute power of the processor is used at $\frac{44}{355/2}
\approx 25\%$ only: a four
times less powerful processor would do the job as well, such as a
monocore at $800$ MHz. In double precision, a $400$ MHz processor
would make it...

From that point, we can take three decisions in order to increase our
computing power:
\begin{itemize}
\item purchase slower processors that use less electricity,
  investing in a larger parallel machine with more compute nodes,
\item give up the low order FDTD scheme and develop high order
  schemes such as Discontinuous Galerkin Method (refer to
  \cite{DGMaxwell} and \cite {Rem02} for the mathematical framework
  and to \cite{DGPRF} for the high order computing benefits shown in
  table $2$ p. $754$) that demand more operations per byte (when
  computing dense-block-matrix products on a Cartesian mesh),
\item reconsider the FDTD implementation on machines with large
  enough data cache.
\end{itemize}
The first point (increasing the number of compute nodes) is investigated in sub-section \ref{cache} and tested in
sub-section \ref{parallel}. On average size problems, it is worth making
parallel computations in a row rather than sequential computations
altogether at once.

The last point, which is the essence of this article, is discussed
section \ref{theory} and tested section \ref{numerexp}, focusing on
techniques that reduce the memory bandwidth requirements through
better using the cache than in the above standard FDTD implementation.

\subsection{Data cache benefits to the FDTD technique}\label{cache}
In the above introduction, we drastically simplified the memory to
processor data access, assuming no cache hierarchy (or problem too
large to take care of cache).
Modern processors no longer possess tiny data caches. We can easily
purchase processors with 2~MB data cache per core (shared or not).
A 2~MB cache can handle a cubic domain made of
$$
\left(\frac{2 \times 1024^2}{4\times 6}\right)^{\frac{1}{3}} = 44
$$
cells per direction in single precision ($4$ bytes per data, $6$
fields per cell). Same formula applied to double
precision gives $35$ cells per direction.
When problem is small enough to fit in processor's data cache, system's
memory is no longer used and we actually benefit from the processor's
increased computing power.

In order to be able to compute larger problems, we can increase the
number of nodes (a way to increase the number of memory controllers to multiply the
available bandwidth) so that the elementary problem on each node fits
in cache.

Let us consider a cubic problem made of $352^3$ cells.
On only one node, data ($4\times 6 \times \frac{352^3}{1024^2} = 998$ MB) does not fit in cache.
Using the $44$ MC/s memory limited
speed with DDR2-800 RAM provided by relation \ref{eq_ram_standard}, we need
$\frac{352^3}{44}\times 10^{-3} = 991$ milliseconds per time step.

When using $512$ nodes, data per node fits in cache
($\frac{352^3}{512} = 44^3$).
In case the work between the nodes is well balanced and assuming a typical
computing speed of $2400$ MFlops per core (i.e. $\frac{2400}{6\times 6}=67$ MC/s), a time step requires
$\frac{36\times 44^3}{2400}\times 10^{-3} = 1.28$ milliseconds.

The limiting factor may become the network's bandwidth for communicating
the inter-domains fields. To parallelize an FDTD solver, one needs to
provide $2$ fields per face on all the interfaces. In a domain with
$6$ neighbours, there are $2 \times 6 \times 44^2 = 23232$ field
values per time step to communicate, i.e. $91$ kB per time step.
On a $10$ Gb/s network (point to point with full duplex), this requires around
$70$ microseconds assuming a perfect communication and computation overlap
(since memory bandwidth is not used for computation).

So, on a perfect network as assumed above, computation time would be more
than $768$ times faster with $512$ nodes: it would take $1.5$ less time to
make $512$ parallel computations on $512$ nodes the one after the
other than to make $512$ sequential computations at the same time.

\subsection{Experimental parallel efficiency superior to one}\label{parallel}
Let us now deal with a real $320^3$ problem spread among $512$
processors (the number of cells is slightly lower to make sure to
actually fit in cache).
On the Tera 10 machine at the CEA's supercomputing centre in Ile de
France, CC-Numa nodes are made of four SMP blocks (called QBB from
Bull) each with independent system bus, each block is made of two $1.6$ GHz
Itanium II dual core processors.
We assume that communication times inside a node are
negligible compared to inter-nodes communication times.
In our example, network connections are to be performed on $32$ nodes.
Let the domain be split so that nodes deal with domains of size
$40\times (\frac{8}{2},\frac{8}{4},\frac{8}{4})$
i.e. $(160,80,80)$ so that the inter-nodes communications are fairly
minimized. This hypothesis can be enforced by an MPI Cartesian
three-dimensional split, or by mixing shared memory programming (to
be handled with care on a Non Uniform Memory Access node) with
distributed memory programming.
\begin{figure}[htbp]
\caption{Minimizing network and NUMA communications: logically identify
  sub-domains to nodes, QBBs, processors and cores.}\label{cartsplit}
\begin{center}
\includegraphics[width=0.7\textwidth]{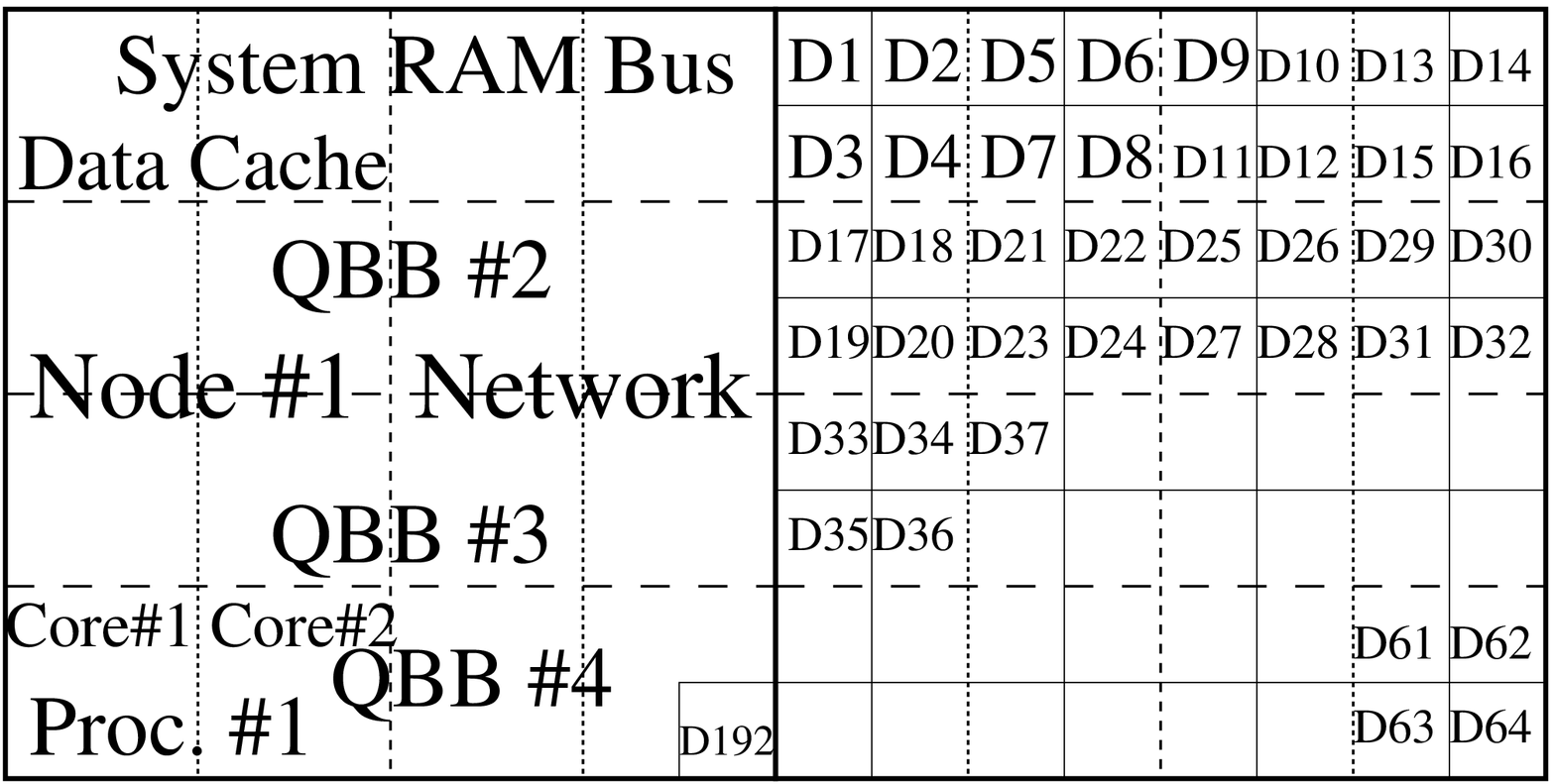}
\end{center}
\end{figure}
Figure \ref{cartsplit} shows on a two-dimensional example how to
map the domains (D) onto the Tera 10 hardware, minimizing inter-nodes
communications first, then intra-nodes communications across the QBB
crossbar interconnect, then intra-processors communications. In the
figure, we put $4$ domains per core to generalize the approach
presented and further discussed in section \ref{theory}.

Using single precision real fields, the number of bytes to
communicate per time step amounts to
$$
4 \times 2 \times \left(80^2+2\times 80 \times 160\right) = 250 \mbox{ kB}.
$$
The network is based on very low latency Quadrics Elan 4 cards, meaning
we can assume a sustainable speed of $900$ MB/s across all nodes.
So, it seems realistic to assume we need
$$
\frac{250}{900\times 1024}\times 10^{3} = 0.27
$$
milliseconds per time step for the communications.
At a peak computing speed of $67$ MC/s per core (lower than the
Itanium's II theoretical speed), we need 
$$
\frac{40^3}{67}\times 10^{3-6} \approx 1
$$
milliseconds per time step for the per core computing time.

There, the parallel efficiency, defined as the time for a computation
made by one core divided by the number of cores $N$ times the time of
the computation with the $N$ cores is at least
$$
\frac{340^3}{44}\times 10^{-3} \times \frac{1}{512\times \left(1+0.27\right)} \approx 1.4
$$
when the communication time is not overlapped by the computation time
at all.

The reader may note that, to manage a computing centre, even though
cores are usually dedicated to only one computation at a time, we may
not reserve the complete processor to a single sequential
application. Thus, in practice, when sequential FDTD computations are
run simultaneously on a multicore processor, the memory bandwidth is
divided by the number of simultaneous sequential computations.

Let us now confront our estimates to an experiment on the Tera 10 machine.
\begin{table}[htbp]
\caption{Practical efficiency on Tera 10 machine, $1000$ time steps.}\label{mpi_sophie}
\begin{center}
\begin{tabular}{|r|r|r|r|r|}
\hline
Cores & Computations & CPU Time(s) & MC/s/core & MPI Wait(s) \cr
\hline
1 & 16 & 2400 & 13.6 & - \cr
1 & 1 & 722 & 45.4 & - \cr
512 & 1 & 1.87 & 34.2 & 0.85\cr
\hline
\end{tabular}
\end{center}
\end{table}
Table \ref{mpi_sophie} shows, on the example of a $320^3$ cube, the
performance that we actually obtained with $32$ nodes of $16$ cores on
Tera 10. First column indicates the number of cores used for the
computation. Second column indicates the number of different
computations performed on the associated nodes. First line stands for
a standard use of the machine, where the sequential code runs alone on
a reserved core, but the other $15$ cores of the node are used for
other numerical simulations. Available bandwidth is shared among $4$
cores and actual performance is reduced by a factor $3.3$ when
comparing with second line where the full node is reserved to only one
busy core. Last column in table \ref{mpi_sophie} provides the MPI
waiting time whereas the third column provides the total CPU time
composed of computation and communication times.

The big disappointing result of this practical test is the sustained
communication time that actually amounts to $3$ times more time than
expected. Another disappointment is that even though fields fit in
cache, communications are not overlapped by computations.

The in cache computing speed is around $2400$ MFlops.
The out of cache speed is around the expected one of $44$ MC/s
when only one core is busy, around $11$ MC/s when all cores are
busy. In this table, we have an apparent efficiency of $2.5$ for the
user who does not use a multicore machine as a monocore one. It takes
two and a half less time to perform $512$ parallel computations on
$512$ cores the one after the other than to make $512$ sequential
computations at the same time. It is then obvious to state that half the
results will be brought back to the user within five times less
time. This advocates that parallel computing can be energy efficient
and very suitable to design optimization where parallelizing on fully
independent problems is paradoxically not always the best technical solution.



\section{Theoretical improvements}\label{theory}
In this section, we present basic ideas to better use data in
cache in order to reduce the stress on the memory bandwidth.
We base our thoughts upon simplified hypotheses that are:
\begin{itemize}
\item processor's data cache to processor's registers throughput is infinite with
no latency,
\item system's RAM latency is fully hidden by prefetching
performed by the compiler.
\end{itemize}
\begin{figure}[htbp]
\caption{Simplified computer system model with infinite Cores to Cache Bandwidth}\label{system}
\begin{center}
\includegraphics[width=0.80\textwidth]{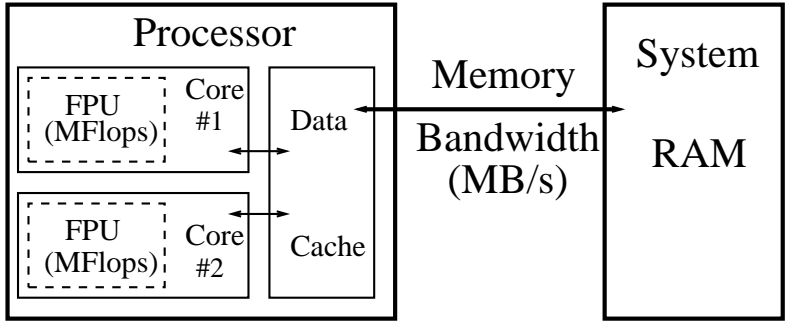}
\end{center}
\end{figure}
Figure \ref{system} shows the simplified data flow and compute engine
models that lie behind our theoretical performance analysis.

In section \ref{multi} we present the idea that consists in emulating
the parallel operating mode in a sequential code, so that elementary
sub-domains handled by the code fit in cache. Thus, the memory
access can be viewed globally on every sub-domain instead of being
performed (with the hypothesis of a very large simulation size) field
by field, leading to a potential halving of the required memory bandwidth.

Section \ref{timescheme} further takes advantage of the sub-domains
decomposition, overlapping the Ampere and Faraday solvers, computing the
global update domain per domain the one after the other instead of
solving the Ampere equation on all the domains, then solving the
Faraday equation on all the domains. Potential gain is another $50\%$.

Under specific assumptions on the sub-domain's physics, we can further
optimize the per domain solver, overlapping the electric and the
magnetic fields update plane by plane as presented in section
\ref{plane}. This technique helps take advantage of the two first ones
even when cache size is reduced.

Then, the final enhancement that consists in solving two or more time
steps together per sub-domain is presented section \ref{cru}. This
idea originates to D. Orozco, PhD student of Pr. Guang R. Gao, University of
Delaware. A potential doubling of the computation speed is expected.
Do note that the first hypothesis no longer applies when memory bandwidth is
optimized with this cache reuse algorithm; a finer analysis should
consider the whole data access hierarchy.

\subsection{Decomposing the problem into smaller problems as in parallel}\label{multi}
As we saw section \ref{parallel}, parallel computations make the problem smaller
per core and induce more cache hits.

So, our idea here is to split the problem into smaller problems, as done
by domain decomposition solvers.
\begin{figure}[htbp]
\caption{Decomposing the large computation domain into many small domains}\label{domains}
\begin{center}
\includegraphics[width=0.60\textwidth]{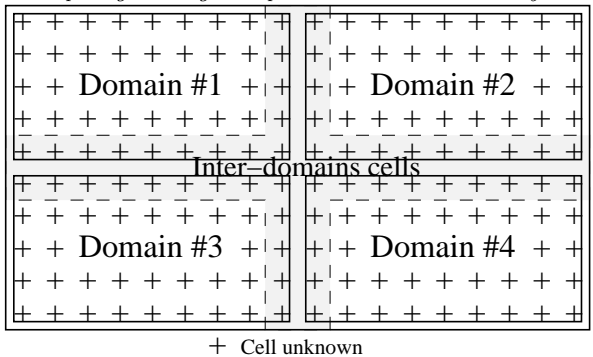}
\end{center}
\end{figure}
Figure \ref{domains} shows an example where we split the computation
domain made of many cells into four domains, each with four times less
cells. We may notice that this decomposition is similar to what can be
naturally performed for parallel processing: introducing a domain
split in a sequential code brings the same programming difficulties as
the ones bound to parallelism. This is a great advantage for designing
a parallel code: most of the bugs will be encountered in the
sequential code where it is usually easier to detect and correct them.

In the FDTD method, the domain decomposition technique is not
a numerical scheme but merely a modified implementation design with no
impact at all on output values, such as loop tiling does. Operations
performed to compute the fields are exactly the same, in the same
order, it is only the order in which we compute the updated values
which differs. This is a general feature for explicit solvers.

So, instead of making triple loops on the whole domain, we make
triple loops in sub-domains, not forgetting to perform double loops at
interfaces to complete the computation.
This technique has already been used in \cite{QSGT} and \cite{QSFD}.

Splitting the domain into sub-domains is done so that every sub-domain
fits in cache, taking care of the actual available cache size for the
multicore case in parallel.

The number of data accesses, when cache conflicts do not occur, is reduced to
\begin{equation}\label{eq_dapf3}
2 \times \frac{3+3+3}{6} = 3
\end{equation}
per field update, instead of $6$ as seen previously in relation
\ref{eq_ram_standard} section \ref{issue}.
Let us explain that.
For the electric field update, we need to load the magnetic
field, the electric field, and then write the electric field.
Without cache conflict, we may compute twice as many
values inside domains as in the monolithic case, but there remains to compute the
values at the interfaces of the domains. The extra-cost becomes negligible if the
sub-domains size is large... which is not possible if the cache size is
too small.
In practice, a 2~MB data cache size is required to really improve performance.

If this article could be read by processors makers, let us ask them
for more cache per core, not only more cores.
In any case, we prefer more cache than higher clock frequencies.

\subsection{Optimizing the leapfrog time scheme implementation}\label{timescheme}
We may go further than merely splitting the problem into
sub-domains. The FDTD time scheme is a leapfrog one where
we solve the Maxwell-Ampere (\ref{ma}) then the Maxwell Faraday (\ref{mf}) equations as follows:
\begin{alltt}
do d=1,ndomains
  call volAmpere(d) ! Computes the internal edges in domain d
enddo
call surfAmpere() ! Computes the edges at domains interfaces
do d=1,ndomains
  call volFaraday(d) ! Computes the internal faces in domain d
enddo
call surfFaraday() ! Computes the faces at domains interfaces
\end{alltt}
\begin{figure}[htbp]
\caption{Internal DoFs to a domain of the leapfrog time scheme}\label{overlap}
\begin{center}
\includegraphics[width=0.5\textwidth]{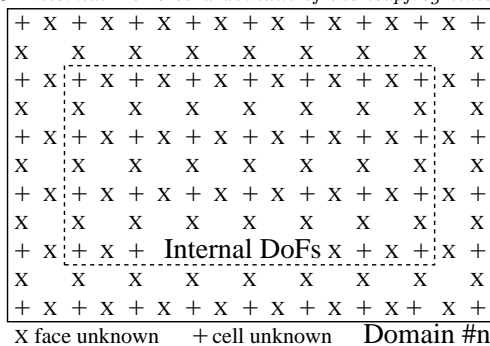}
\end{center}
\end{figure}
The improvement consists in solving Ampere and Faraday in a row using
the internal degrees of freedom (DoFs):
\begin{alltt}
do d=1,ndomains
  call volAmpere(d)
  call volFaraday(d) ! Internal faces that use internal edges only
enddo
call surfAmpere()
call surfFaraday() ! Adds computation at faces using edges on interfaces.
\end{alltt}

Figure \ref{overlap} shows in a two-dimensional space the internal degrees of
freedom that are updated with the modified time scheme: the electrical
field is located at cells (``$+$'') and the magnetic field is located
at faces (``$\times$''). We no longer update the magnetic field values
that need an electric value at the sub-domain's boundary. This
technique can be applied to the heat diffusion equation, not only to
the waves equation.

With the inter-leaved scheme, assuming there is no cache conflict, we perform
\begin{equation}\label{eq_dapf2}
2 \times \frac{3+3}{6} = 2
\end{equation}
memory accesses per field update: we load the electric and the magnetic
fields, then we write them. This number is to be compared to the data
access count of relation \ref{eq_dapf3} section \ref{multi}.

As compared with the initial standard FDTD implementation (relation
\ref{eq_ram_standard} section \ref{issue}), the
computation speed we forecast is
\begin{equation}\label{eq_ram_m6}
\frac{6400}{4\times 6\times 2} = 133 \mbox{ MC/s}.
\end{equation}

So, slightly altering the time scheme implementation, after
decomposing the problem into sub-problems, leads to a theoretical
$50\%$ increase in computing efficiency.
When designing a parallel code, this brings a strong impact on the
code's architecture since no communication at all must be done between the
Ampere and the Faraday updates on the contrary to what is generally
performed such as in \cite{MPIFDTD} or \cite{GUIF}.

Making only one communication per time step can also be
profitable to the code's parallel performance avoiding an inconvenient
superfluous network latency.

\subsection{Overlapping the leapfrog time scheme plane by plane}\label{plane}
In an FDTD code, the basic vacuum equations may not be the only ones
to solve between the electric field update and the magnetic field
update. To take account of boundary conditions, for instance on a
PEC (Perfectly Electric Conductor) we need to zero the electric field
values on the PEC before entering the Faraday solver.
In materials, we may need to take into account some recursive
accumulators as in the JEC technique (see \cite{JEC}) for linear
dispersive media, we may need an absorbing boundary layer to emulate
the infinite radiation condition (see \cite{CPML}), or simply need to
add a current to Ampere's equation. Materials and PML layers
optimization is not dealt with in this article for the sake of
conciseness, but basing the solver on one-dimensional geometrical
coefficients as presented section \ref{issue} instead of easier to implement
three-dimensional ones is essential not to increase the memory
bandwidth stress.

We consider domains where there is nothing else to do but solve the
Maxwell equations in vacuum.
This assumption is not useless when the problem is split in many
sub-domains, since there probably will exist many sub-domains where the assumption
will be true. The reader may notice that this holds for a parallel code,
but the gain shall be limited by the difficulty to make larger or
smaller domains according to the associated processing speed. When
using many sub-domains, even in a parallel computation, we may expect
the load balancing to be naturally correct, especially if we layout
the domains onto the processors using a balanced cost function.

So, in that vacuum case, the electric field and the magnetic field
updates can be done plane by plane along with index k representing the z axis
discretization.
In the case of a 2~MB cache in single precision, this technique provides
a good cache reuse for cubic domains of size up to
$$
\left(\frac{2 \times 1024^2}{4\times (3+3+3+3)}\right)^{\frac{1}{2}} = 209
$$
cells per direction, which is far more than the value of $44$ found in
section \ref{cache}. Moreover, this is independent of the $z$
direction size.

The algorithm becomes:
\begin{alltt}
do d=1,ndomains
  call planeAmpere(d,1)
  do k=2, nz
    call planeAmpere(d,k); call planeFaraday(d,k-1)
  enddo
  call planeFaraday(d,nz)
enddo
call surfAmpere(); call surfFaraday()
\end{alltt}
Of course, programming becomes rather more complicated than the
one presented in Taflove's handbook \cite{Taflove}. This may become
a drawback for the compiler's ability to perform optimizations.
In particular, compiler-assisted cache replacement \cite{GaoCache2} may be less
efficient.

\subsection{Getting rid of the memory bandwidth bottleneck with limited cache stress}\label{cru}
In vacuum, with the same assumption as in the above section, we may
get rid of the memory bandwidth bottleneck using a very simple idea
from Daniel Orozco \cite{Orozco}.

\begin{figure}[htbp]
\caption{Internal DoFs to a domain with the two steps cache reuse algorithm}\label{crudofs}
\begin{center}
\includegraphics[width=0.5\textwidth]{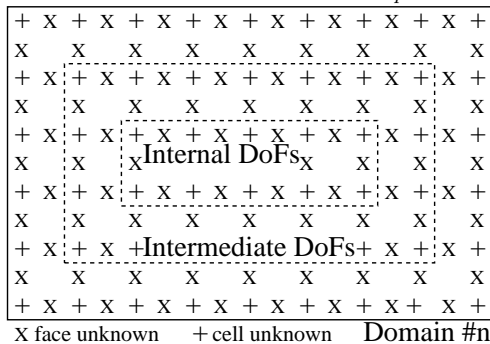}
\end{center}
\end{figure}

Instead of performing the sub-domains updates step by step, we
gather time steps. For the start, we use two steps as given by
the following pseudo-code:
\begin{alltt}
do n=1,ntimesteps/2
  ! Compute first and second time steps altogether inside domains:
  do d=1,ndomains
    call planeAmpere(d,1,nx,ny)
    call planeAmpere(d,2,nx,ny)
    call planeFaraday(d,1,nx,ny)
    ! Compute the internal DoFs by plane for two time steps in a row
    do k=3, nz
      call planeAmpere(d,k,nx,ny)
      call planeFaraday(d,k-1,nx,ny)
      call planeAmpere(d,k-1,nx-1,ny-1)
      call planeFaraday(d,k-2,nx-1,ny-1)
    enddo
    call planeFaraday(d,nz,nx,ny)
  enddo
  ! Terminate the first time step:
  call surfAmpere(); call surfFaraday()
  ! Terminate the missing volume computations for the second time step:
  do d=1,ndomains
    call surfAmpere(d,nx-1,ny-1,nz-1)
    call surfFaraday(d,nx-1,ny-1,nz-1)
  enddo
  ! Terminate the second time step:
  call surfAmpere(); call surfFaraday()
enddo
\end{alltt}
Figure \ref{crudofs} shows the very internal DoFs that are computed over
two time steps every even time step and the intermediate internal
DoFs that need be computed at every time step. These intermediate
DoFs form a surface layer of one cell width next to the sub-domain's boundary.

If we neglect the cost of terminating the computation for the internal
DoFs of the domains near the boundaries, we use half the
memory bandwidth as long as the double plane problem fits in cache,
thus for cubic domains of size up to
$$
\left(\frac{2 \times 1024^2}{4\times (3\times 6)}\right)^{\frac{1}{2}} = 170
$$
cells per direction.

Doing that, assuming there is no cache conflict, we need
only one memory access per field update, theoretically doubling the
speed of the FDTD code.

Theoretical performance due to memory bandwidth in case of cache
fitting and under the assumption there is no cache conflict thus
should reach:
\begin{equation}\label{eq_ram_m12}
\frac{6400}{4\times 6\times 1} = 267
\end{equation}
million cells per second. At this speed level, assuming cache to
processor's throughput is infinite no longer holds.

The drawbacks of this technique are of course a very complicated
programming which may limit the compiler's efficiency, but also
 merely outputting a field at a pin point becomes
tricky to handle. So, instead of not performing this optimization, the right way
to do when a pin point value is requested at a time step, is to avoid
starting the two time steps the time step before the output is needed. When
dealing with a computation over hundreds of thousands time steps,
the user shall rarely ask for a print every time step.


\section{Numerical experiments}\label{numerexp}
We make numerical experiments benchmarking the various optimizations
suggested section \ref{theory}. Section \ref{perfma} tests the
kernel's speed on sub-domains, not taking the inter-domains over-cost
into account. Section \ref{sophie} presents two complete simulation
tests and the obtained overall gains on the production's code Sophie
that mixes up the different algorithms of section \ref{theory}.

\subsection{Experimental performance analysis}\label{perfma}
In this section, we analyze performance results in million cells
computed per second (MC/s) obtained on an Intel Q6600 quadcore processor at
$2.4$ GHz with two 4~MB cache shared by pair.
In monocore mode, the available cache shall be 4~MB, in quadcore mode
it will decrease down to 2~MB. In this article, we shall not discuss
the effects of shared or dedicated cache.

The memory is made of four 2 GB DDR2-1066 dual channel modules.
We investigate, for single or double precision, for monocore or
quadcore being active, the performance variations along with the
problem's size $N$, $N$ being the equal number of cells per direction.
Computer implementation is done in FORTRAN and compiled
using ``Intel 10.1.018 FORTRAN compiler for 64 bits applications''.

The operating system is Mandriva Spring $2008.1$ with Linux $64$ bits
kernel $2.6.24$ and MPI is OpenMPI release $1.2$.

\begin{figure}[htbp]
\caption{Standard FDTD programming, multicore of limited interest}\label{perfma_m4}
\includegraphics[width=0.95\textwidth]{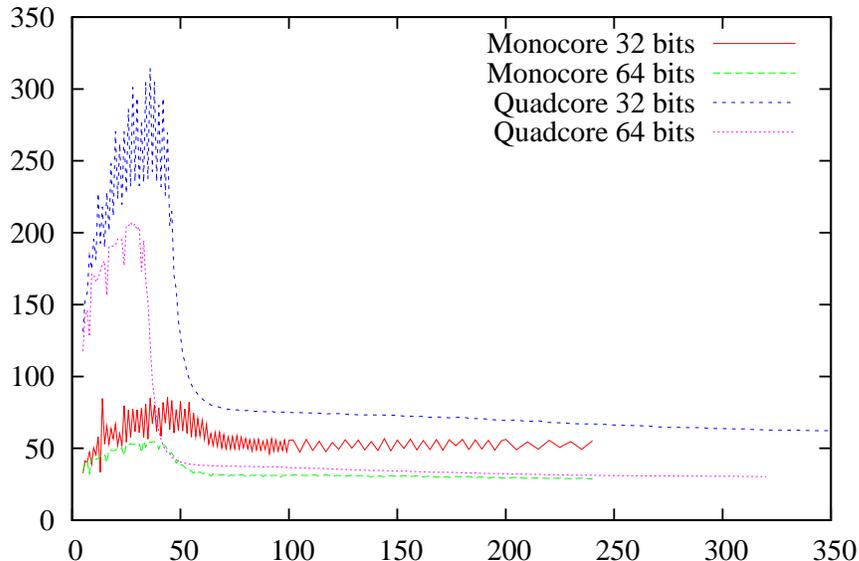}
\end{figure}
Figure \ref{perfma_m4} indicates that performance per core in single precision
is close to $2400$ MFlops ($67$ MC/s), the theoretical performance peak of a $2.4$
GHz Core 2 processor family. This performance is sustained oscillating
around $1800$ MFlops in single precision with only one core being
busy.
When all cores are busy, performance exceeds the $4 \times 2400$
MFlops peak processor's speed ($267$ MC/s), which is a measuring
artefact. Performance drops for problems of size $44$ as expected
since data no longer fit in cache.
When problem's size becomes larger and larger, the global
performance of the four cores working altogether or only one core being
active get closer and closer. This means that standard FDTD
codes do not benefit from multicore, apart from the very
academic cases where the simulation domain is small enough to fit in
the processor's data cache.
In single precision computing, the asymptotic performance value for large
domains drops down to below $62$ MC/s (million cells per second).
In double precision, the asymptotic value seems to be reached at around $30$
MC/s.
The theoretical value, on single precision code, is given by equation \ref{eq_ram_standard} for a
$1066$ MHz RAM:
$$
\frac{1066}{800} \times\frac{6400}{4\times 6\times 6} = 59 \mbox{ MC/s}.
$$
This is very close to the value provided by the numerical experiment.

\begin{figure}[htbp]
\caption{Splitting into sub-domains, multicore rather interesting}\label{perfma_m6}
\includegraphics[width=0.95\textwidth]{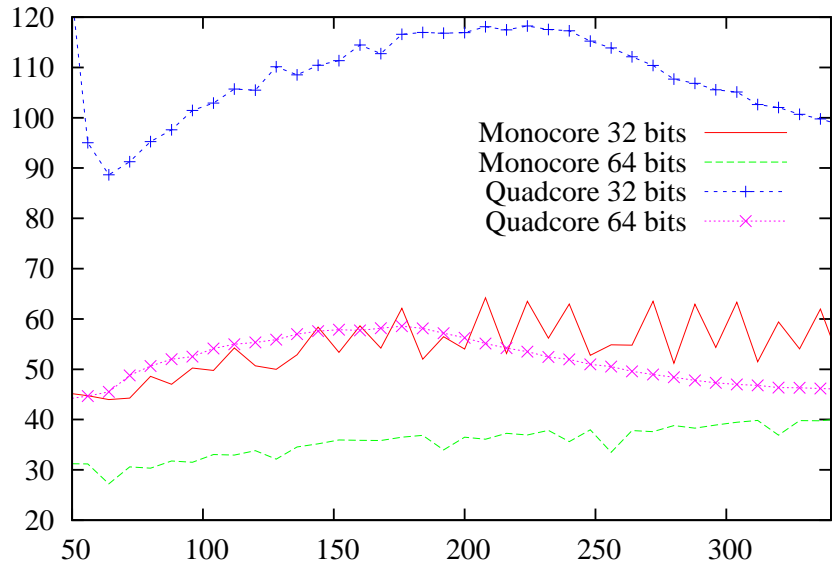}
\end{figure}
Figure \ref{perfma_m6} shows the benefit of splitting the problem into
sub-domains (section \ref{multi}) using the modified time scheme (section \ref{timescheme}).
In this figure, instead of
using one large problem of size $N$, we use $512$ domains of size $N/8$
per axis. The figure indicates the global problem size $N$. The
problem, as decomposed, fits in cache for very small sub-domains, so we
start at $N=48$. We observe that the multicore is now useful,
bringing a doubling of the computation speed even for rather
large problems. The optimal sub-domains size is around $27$ in single
precision and $22$ in double precision to fully benefit from
multicore. This value, lower than the expected value of $44$ is
probably due to cache conflicts.
When using adequate sub-domains size, we get a throughput of around $120$
MC/s (million cells per second) in single precision. This is twice the
speed of the standard FDTD implementation, but when using equation
\ref{eq_ram_m6} of section \ref{timescheme}, we expect a
throughput of
$$
\frac{1066}{800} \times\frac{6400}{4\times 6\times 2} = 178 \mbox{ MC/s}
$$
instead of $120$. Instead of a factor $3$ that could be theoretically
expected, we only get a factor $2$.
In double precision, we get a speed at around $58$ MC/s instead of $30$
with the standard FDTD implementation. In the monocore mode, speed
increases with the problem's size up to $40$ MC/s which shows that the
data access pattern is improved. The double precision code speed is
tightly controlled by the memory bandwidth and thus better benefits
from this kind of loop tiling.

\begin{figure}[htbp]
\caption{Plane solver on sub-domains, enlarging the range for multicore interest}\label{perfma_m10}
\includegraphics[width=0.95\textwidth]{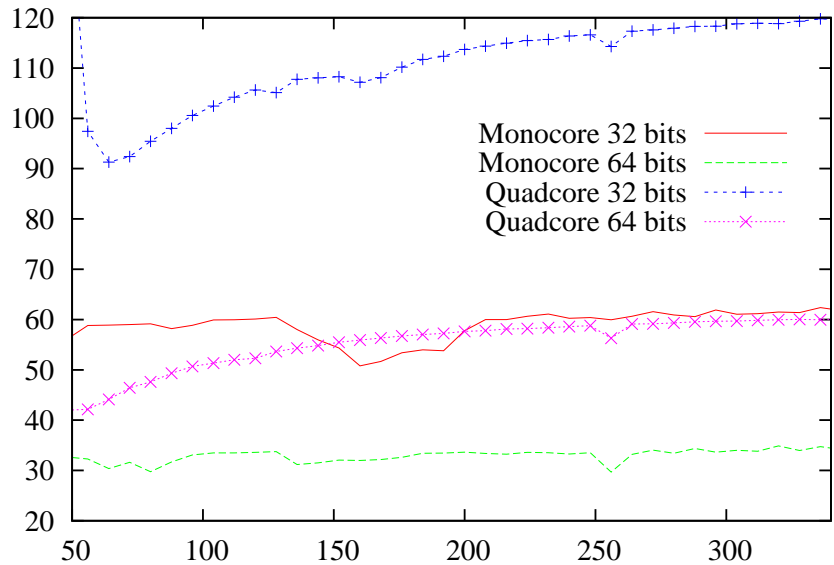}
\end{figure}
Figure \ref{perfma_m10} shows the interest of solving the Ampere and
Faraday equations together plane by plane (section \ref{plane}). This leads to less cache
conflicts and performance increases beyond sub-domains of size
$50$ in multicore. In monocore, performance is very stable.
So, the interest of this technique is not to increase the computation
speed related to the performance observed in the previous figure~\ref{perfma_m6},
but to make this performance sustainable on a larger
domain size range, especially useful if the cache size per core were
to decrease in the future with ever growing number of cores but not
necessarily growing data cache size per core. Let us note that the compiler did a
good job since the monocore 32 bits performance curve is maintained at
the previous level of figure \ref{perfma_m6}. The double precision
monocore speed is lower than before, but not significantly.

\begin{figure}[htbp]
\caption{Cache reuse, quadcore two and a half as fast as monocore}\label{perfma_m12}
\includegraphics[width=0.95\textwidth]{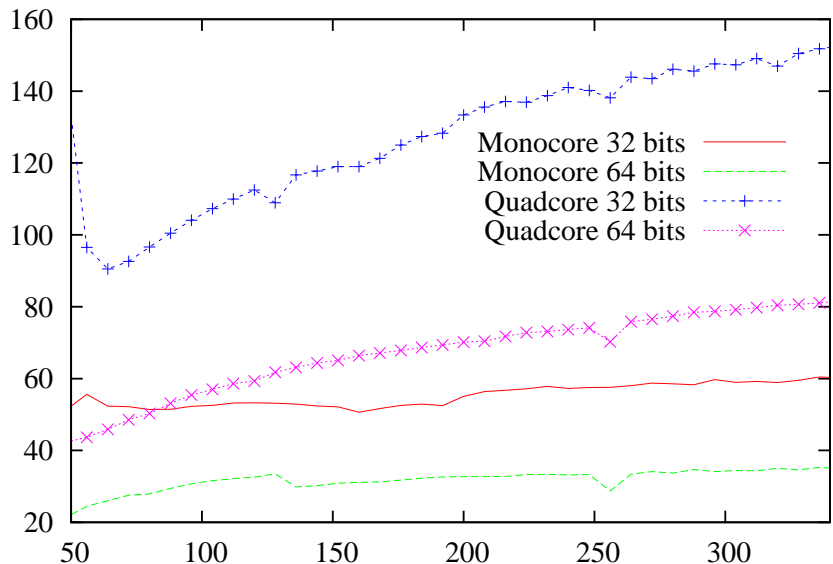}
\end{figure}
Figure \ref{perfma_m12} shows the tremendous advantage of the cache reusing technique
solving two time steps at once (section \ref{cru}). Multicore performance increases with
the problem's size, optimal sub-domains size exceeding the value $N=50$.
We obtain a performance gain in multicore of more than two and a half
as compared with the monocore performance, indicating that we are on
the way to get rid of the memory bandwidth constraint.
Nevertheless, in the monocore mode, this technique does not bring much benefit for small
domains: the source code becomes difficult to optimize for the
compiler and there is no benefit unless the bottleneck is actually the
memory bandwidth.
Performance on quadcore peaks at $160$ MC/s in single precision  and
$80$ MC/s in double precision.
When using formula \ref{eq_ram_m12} with DDR2-1066, we ought to get a
single precision speed of
$$
\frac{1066}{800} \times\frac{6400}{4\times 6\times 1} = 355 \mbox{ MC/s}
$$
which is more than twice the experimental speed of figure
\ref{perfma_m12}. But there, the limiting factor would no longer be
the memory bandwidth, since the peak processor's performance, being
four times $2400$ MFlops, limits the computation throughput to a
maximum of
$$
4 \times \frac{2400}{36} = 266  \mbox{ MC/s.}
$$

Thus, the practical performance drops only by a $1.6$ factor.
This can be justified by cache conflicts and reduced
code optimization due to compiler's failure to perfectly optimize
a much more complicated source code.
Nevertheless, when comparing with the standard FDTD code of
figure \ref{perfma_m4}, we obtain a $2.7$ gain in computation speed,
both in double and in single precision.
This explains why this article in entitled ``getting rid of the memory
bandwidth bottleneck'' with an apparent equivalent bandwidth of
$$
\frac{160}{59} \times 6400 \times \frac{1066}{800} = 22.5 \mbox{ GB/s}.
$$

\subsection{Global performance observed on real simulations using Sophie}\label{sophie}
The Sophie code is a house made CEA/DAM software for
electro-magnetics simulations solving Yee's \cite{Yee} or Yee's dual
scheme (inverting the locations of the fields between edges and faces)
with an FDTD solver. The code's first developments date back to mid 2005
and production's first release dates back to end 2006. It is a general
purpose solver that can model any linear dispersive medium (according
to \cite{JEC} with slight unpublished improvements), represent
the infinite radiation condition (from \cite{CPML} with an
implementation that minimizes memory bandwidth's demand),
approximate thin layers by impedance conditions (following
\cite{SUMERT}), approximate thin wires (using Holland's model
\cite{Fil} improved in \cite{Filb}) and compute their coupling with
simple electronic circuits. A basis for Sophie's implementation is
found in C. Guiffaut's PhD Thesis \cite{GUIF} with further improvements orally
provided by it's author.

Sophie is a massively parallel code designed to run on thousands of
multicore SMP nodes targeting tera-cells computations by the end of the
decade on the Tera 100 machine.
Sophie is made of 200 thousand lines of FORTRAN, most of them
being related to the multi-domain feature initially brought in for both
multi-physics load balancing and easier parallel debugging. A key
feature for easing maintenance is that parallel and sequential runs
provide exactly the same output, with no machine round-off difference.
This tremendously helps eliminating the parallel bugs, making
automatic non-regression parallel tests prove the code's validity on a
huge number of parallel configurations. Sophie is accompanied by an
incredible ten thousand non-regression parallel tests that graph proof
the code's integrity.

For this article, we considered two cavity test cases:
\begin{enumerate}
\item a sphere included in a domain of size $402^3$ with a diameter
  made of $400$ cells,
\item a cubic domain of size $402^3$ with no internal object.
\end{enumerate}
In both cases, the problem amounts to around $64$ million cells, and
we perform $1300$ time steps, i.e. $108\%$ of the domain's diagonal.
We set a current at a pin point inside the sphere, we choose the same
pin point for the cube. We also output 6 points at various positions
every 20 time steps (not too often, in order not to stop the cache
reuse algorithm too many times).
We make various domain's splits to see the impact on global
performance.

\begin{table}[htbp]
\caption{Sophie production code's performance table, 64 bits.}\label{perfs_sophie}
\begin{center}
\begin{tabular}{|c|r|c|c|r|r|r|}
\hline
Case & Split & Cores & MC/s & Time (s) & Vol. Time (s) & V/T \% \cr
\hline
S    & 1     & 1     & 27   & 3160 & 3160 & 100 \cr
S    & 1,2,2 & 4     & 30   & 2836 & 2836 & 100 \cr
S    & A     & 1     & 38   &  1927 &  1303 & 72 \cr
S    & A     & 4     & 58   &  1264 &  776 &  66 \cr
S    & P2    & 4     & 58   &  1298 &  661 & 56 \cr
S    & P3    & 4     & 37   &  2075 &  1063 & 57 \cr
S    & P4    & 4     & 55   &  1314 &  919 & 75 \cr
S    & P5    & 4     & 56   &  1296 &  877 & 73 \cr
\hline
C    & 1     & 1     & 27   & 3103 & 3103 & 100 \cr
C    & 1,2,2 & 4     & 30   & 2810 & 2810 & 100 \cr
C    & A     & 1     & 31   &  2292 &  1668 & 77 \cr
C    & A     & 4     & 59   &  1227 &  765 & 66 \cr
C    & P2    & 4     & 49   &  1486 &  861 & 63 \cr
C    & P3    & 4     & 37   &  2020 &  1011 & 54 \cr
C    & P4    & 4     & 67   &  1082 &  702 & 71 \cr
C    & P5    & 4     & 67   &  1080 &  683 & 69 \cr
\hline
\end{tabular}
\end{center}
\end{table}

Table \ref{perfs_sophie}, for double precision computation, shows for the sphere (S) and cube (C) cases
the performance observed at the last time step in million cells per
seconds (MC/s), as well as the total elapsed time in seconds (over all the
time steps).

These numbers are not linearly deducible the one from the
other except for the mono-domain case. In the case of sub-domains, we
do not perform computations on sub-domains with all zeroed fields on
their boundary as in the plane case of section \ref{plane}.
This is a rather trivial optimization, which is particularly interesting at
the beginning of the simulation. Do note that for the sphere case, if
the sub-domains are small enough, there will always remain some
domains entirely with zeroed fields.

After the total CPU time column, the following column in table
\ref{perfs_sophie} indicates the total CPU time spent in the
sub-domains volume update routines. In multi-domain, a large part of the CPU time may be
spent in the inter-domains routines \verb+surfAmpere+ and
\verb+surfFaraday+ (as seen in section \ref{timescheme}).
Last column indicates, at the last time step, the percentage ratio of
the time spent in the volume computations to the total CPU time.

Performance varies according to the number of cores we
use, and, above all, according to the problem's split that is
performed.
The $(1,2,2)$ split builds four domains cut in the y and z planes.
The ``A'' split is the automatic split in $(6,16,16)$ performed by the code, as a
result of the study of the preceding section.
The ``P2'' split is a manual dense split in $(10,24,12)$.
The ``P3'' split is a manual dense split in $(2,40,20)$ that favours the
x axis length, which is supposed to be good for vectorization and thus performance.
The ``P4'' split is a manual standard symmetric split in $(8,8,8)$.
The ``P5'' split is a manual coarse split in $(6,12,8)$.

Since the cache reuse technique cannot be
performed on the sphere's PEC (Perfectly Electric Conductor) boundary, a coarse split limits the
number of cells that are efficiently performed. Furthermore, too large
domains are not suitable for fitting in cache. This is why the P5
split is not suitable for the sphere. The P3 split inhibits the cache reuse algorithm on all domains and is even worse.

Nevertheless, we have seen section \ref{perfma} that larger domains are computed faster with the cache reusing
technique. Thus, for the cube case where all sub-domains can benefit from that
programming, we obtain a $2.3$ times speed increase to the initial
standard FDTD code in parallel.
This follows the performance laws presented in section \ref{perfma} for the inside volumes computations.

\begin{table}[htbp]
\caption{Sophie production code's performance table, 32 bits.}\label{perfs6_sophie}
\begin{center}
\begin{tabular}{|c|r|c|c|r|r|r|}
\hline
Case & Split & Cores & MC/s & Time (s) & Vol. Time (s) & V/T \% \cr
\hline
S    & 1     & 1     & 43   & 1953 & 1953 & 100 \cr
S    & 1,2,2 & 4     & 62   & 1376 & 1376 & 100 \cr
S    & A     & 1     & 61   &  1183 &  817 & 74 \cr
S    & A     & 4     & 112   &  658 &  367 &  62 \cr
S    & P2    & 4     & 102   &  724 &  342 & 51 \cr
S    & P3    & 4     & 76   &  990 &  468 & 51 \cr
S    & P4    & 4     & 105   &  694 &  450 & 72 \cr
S    & P5    & 4     & 107   &  679 &  425 & 69 \cr
\hline
C    & 1     & 1     & 44   & 1905 & 1905 & 100 \cr
C    & 1,2,2 & 4     & 62   & 1358 & 1358 & 100 \cr
C    & A     & 1     & 51   &  1376 &  1012 & 78 \cr
C    & A     & 4     & 109   &  670 &  380 & 64 \cr
C    & P2    & 4     & 85   &  852 &  456 & 60 \cr
C    & P3    & 4     & 73   &  1023 &  503 & 54 \cr
C    & P4    & 4     & 118   &  615 &  371 & 67 \cr
C    & P5    & 4     & 120   &  602 &  364 & 67 \cr
\hline
\end{tabular}
\end{center}
\end{table}
Table \ref{perfs6_sophie} is identical to table \ref{perfs_sophie}
for single precision computation. Same remarks can be made, with
slightly smaller gains since the memory bandwidth is less the major limiting factor than in double precision.
The performance is nevertheless nearly doubled.

\section{Conclusion and outlooks}
We have dealt with the performance bottleneck that
limits the computing efficiency of FDTD codes. Using ideas borrowed to
the computer scientists community (in particular discussing with
D. Orozco), we have exhibited a technique to
start getting rid with the memory constraint on multicore processors
with large data caches. This technique has been implemented in a
production code where it demonstrates a gain by a factor two on
real applications.


A perspective to this cache reuse technique could be to gather more and more
time steps together. The practical gain would decrease as a function
of the sub-domains size, making high number of time steps gathering
inefficient. For a limited number of time steps, say for instance $6$,
the difficulty (apart from programming the $6$ elementary update
functions) would be to manage the pin points prints and the fields
images prints. For an industrial code that must be able to cope with
6 steps gathering as well as $2$ (for smaller domains), mixing the steps
would be difficult. In a multi-physics code, the interest might also be
reduced.

Even though we limited ourselves to an FDTD code, this cache reuse technique can be
applied to any low order solver on Cartesian meshes. It can also be applied to
solvers based upon free meshes, despite it's efficiency shall be more limited
by the available cache size since fields indirect addresses and sparse
update matrix coefficients must also fit in cache, not only the fields values.

Last, but not least, conceiving applications keeping in mind the cache
to RAM data transfers is suitable to general purpose GPU
implementation where (from a high level software's architecture point
of view) GPU's RAM plays the role of the processor's cache
and where the limiting factor is the memory transfers to system's RAM
through the PCI bus.

\section*{Acknowledgments}
The author thanks Pierre Lalet, computer science engineer at CEA,
for invaluable remarks and suggestions on this work.

\addcontentsline{toc}{section}{Bibliography.}
\bibliography{biblio}

\begin{thebibliography}{10}

\bibitem{Gems}
{\sc Ulf Anderson}, {\em Yee bench, a {PDC} benchmark code}, University of
  Stockholm,  (2002).

\bibitem{DGPRF}
{\sc M.~Bernacki, L.~Fezoui, S.~Lanteri, and S.~Piperno}, {\em Parallel
  discontinuous {G}alerkin unstructured mesh solvers for the calculation of
  three-dimensional wave propagation problems}, Applied Mathematical Modelling,
  30 (2006), pp.~744--763.

\bibitem{HMPP2}
{\sc F.~Bodin and S.~Bihan}, {\em Heterogeneous multicore parallel programming
  for graphics processing units}, Scientific Programming Journal,  (to appear
  in).

\bibitem{MC96}
{\sc M.~Cessenat}, {\em Mathematical Methods in Electromagnetism}, vol.~41 of
  Series on Advances in Mathematics and Applied Sciences, World Scientific,
  1996.

\bibitem{JEC}
{\sc Q.~Chen, M.~Katsurai, and P.~Aoyagi}, {\em An {FDTD} {F}ormulation for
  dispersive media using a current density}, IEEE Transactions on Antennas and
  Propagation, 46 (1998), pp.~1739--1745.

\bibitem{Filb}
{\sc Francis Collino and Florence Millot}, {\em Fils et m{\'e}thodes
  d'{\'e}l{\'e}ments finis pour les {\'e}quations de {M}axwell. {L}e mod{\`e}le
  de holland revisit{\'e}.}, Tech. Report 3472, INRIA-Roquencourt, CERFACS, 42
  avenue G. Coriolis, F-31057 Toulouse, August 1998.

\bibitem{Ding}
{\sc C.~Ding}, {\em Improving Effective Bandwidth through Compiler Enhancement
  of Global and Dynamic Cache Reuse}, PhD thesis, Rice University, Texas, 01
  2000.

\bibitem{DingKen}
{\sc Chen Ding and Ken Kennedy}, {\em The memory bandwidth bottleneck and it's
  amelioration by a compiler}, in International Parallel and Distributed
  Processing Symposium, Cancun, Mexico, 05 2000.

\bibitem{CPML}
{\sc S.~Gedney}, {\em Convolution {PML} ({CPML}): An efficient {FDTD}
  implementation of the {CFS-PML} for arbitrary media}, Microwave and Optical
  Technology Letters, 27 (2000), pp.~334--339.

\bibitem{GR00}
{\sc Stephen~D. Gedney and J.~Alan Roden}, {\em Numerical stability of
  nonorthogonal {FDTD} methods}, IEEE Transactions on Antennas and Propagation,
  48 (2000), pp.~231--239.

\bibitem{TUM1}
{\sc F.~Guenther, M.~Mehl, M.~Poegl, and C.~Zenger}, {\em A cahe-aware
  algorithm for {PDE}s on hierarchical data structures based on space-filling
  curves}, SIAM J. Sci. Computing, 28 (2006), pp.~1634--1650.

\bibitem{GUIF}
{\sc Christophe Guiffaut}, {\em Contribution {\`a} la m{\'e}thode {FDTD} pour
  l'{\'e}tude d'antennes et de la diffraction d'objets enfouis.}, PhD thesis,
  FRANCE/Universit{\'e} de Rennes I, Traitement du Signal et
  T{\'e}l{\'e}communications., 10 2000.

\bibitem{GaoCache2}
{\sc G.~R.~Gao H.~Yang, R.~Govindarajan and Z.~Hu}, {\em Compiler-assisted
  cache replacement: Problem formulation and performance evaluation}, in 16th
  International Workshop on Languages and Compilers for Parallel Computin,
  College Station, Texas, 10 2003, LCPC'03.

\bibitem{TUM2}
{\sc J.~Hartmann, A.~Krahnke, and C.~Zenger}, {\em Cache efficient data
  structures and algorithms for adaptive multidimensional multilevel finite
  element solvers}, Applied Numerical Mathematics, 58 (2008), pp.~435--448.

\bibitem{DGMaxwell}
{\sc J.~S. Hesthaven and T.~Warburton}, {\em Nodal high-order methods on
  unstructured grids - time-domain solution of {M}axwell's equations}, Journal
  of Computational Physics, 181 (2002), pp.~186--221.

\bibitem{Fil}
{\sc Richard Holland and Larry Simpson}, {\em Finite-difference analysis of emp
  coupling to thin struts and wires}, IEEE Transactions on Electromagnetic
  Compatibility, 23 (1981), pp.~88--97.
\newblock Mission Research Corporation, Albuquerque USA.

\bibitem{ComGems}
{\sc R.~Mittra, Wenhua Yu, Yongquan Lu, and Rui Lu}, {\em Gems - a general
  purpose conformal {FDTD} solver tailored for parallel platforms}, in
  Asia-Pacific Symposium on Electromagnetic Compatibility, 05 2008.

\bibitem{Orozco}
{\sc D.~Orozco}, {\em Gathering time steps per domain in order to better reuse
  data in cache}.
\newblock Private communication, 07 2008.

\bibitem{Rem02}
{\sc S.~Piperno, M.~Remaki, and L~Fezoui}, {\em A non-diffusive finite volume
  scheme for the three-dimensional {M}axwell's equations on unstructured
  meshes}, SIAM Journal, Numerical Analysis, 39 (2002), pp.~2089--2108.
\newblock INRIA, Sophia Antipolis.

\bibitem{SUMERT}
{\sc C.~Le Potier}, {\em Le code {S}umer-{T}, calcul d'\'electromagn\'etisme
  3{D} temporel par une formulation volumes finis avec condition aux limites
  absorbante}, Tech. Report DO91, CEA, CEL-V, 1993.

\bibitem{GaoRegister}
{\sc Hongbo Rong, Alban Douillet, and Guang~R. Gao}, {\em Register allocation
  for software pipelined multi-dimensional loops}, in 2005 ACM SIGPLAN
  conference on Programming language design and implementation, Chicago, IL,
  USA, 2005, pp.~154--167.

\bibitem{CUDA}
{\sc S.~Ryoo, C.~I. Rodrigues, S.~Baghsorkhi, S.~Stone, D.~Kirk, and W.~Hwu},
  {\em Optimization principles and application performance evaluation of a
  multithreaded {GPU} using {CUDA}}, in 13th ACM SIGPLAN Symposium on
  Principles and practice of parallel programming, Salt Lake City, 2008.

\bibitem{QSGT}
{\sc D.~Seidel, R.~Coats, W.~Johnson, M.~Kiefer, P.~Mix M., Pasik~T. Pointon,
  J.~Quintenz D., and Riley~D. Turner}, {\em Quicksilver, a general tool for
  electromagnetic {PIC} simulation}, tech. report, Sandia National
  Laboratories, Albuquerque, New Mexico 87175, 1996.

\bibitem{QSFD}
{\sc D.~Seidel, R.~Coats, M.~Kiefer, T.~Pointon, and J.~Quintenz}, {\em
  Quicksilver, a 3d, time-domain, finite-difference code for the
  electromagnetic simulation of complex structures}, Tech. Report DE92-003147,
  Sandia National Laboratories, Albuquerque, New Mexico 87175, 1992.

\bibitem{MPIFDTD}
{\sc M.~Su, D.~Bader, I.~El-Kady, and S.-Y. Lin}, {\em A novel {FDTD}
  application featuring {OpenMP-MPI} hybrid parallelization}, in 2004
  International Conference on Parallel Processing, ICPP04, 2004.

\bibitem{Taflove}
{\sc Allen Taflove}, {\em Computational {E}lectrodynamics, {T}he
  {F}inite-{D}ifference {T}ime-{D}omain {M}ethod}, Artech House Publishers,
  1995.

\bibitem{Cell}
{\sc L.~Verducci, P.~Placidi, P.~Ciampolini, A.~Scorzoni, and L.~Roselli}, {\em
  A standard cell hardware implementation for finite-difference time domain
  ({FDTD}) calculation}, IEEE Transactions on Microwave Theory and Techniques,
  (2003).

\bibitem{HMPP1}
{\sc Workhop on General Processing Using GPUs}, {\em {HMPP}: A Hybrid
  Multi-core Parallel Programming Environment}, Boston, 10 2007.

\bibitem{GaoCache1}
{\sc Hongbo Yang, R.~Govindarajan, G.~Gao, and Ziang Hu}, {\em Improving power
  efficiency with compiler-assisted cache replacement}, Journal of Embedded
  Computing, 1 (2005), pp.~487--499.

\bibitem{Yee}
{\sc Kane~S. Yee}, {\em Numerical solution of initial boundary value problems
  involving {M}axwell's equations in isotropic media}, IEEE Transactions on
  Antennas and Propagation, 14 (1966), pp.~302--307.

\end{thebibliography}
\end{document}